\documentclass[11pt]{article}

\usepackage[preprint]{acl}

\usepackage{times}
\usepackage{latexsym}

\usepackage[T1]{fontenc}

\usepackage[utf8]{inputenc}

\usepackage{microtype}
\usepackage{algorithm}  
\usepackage{algorithmicx}  
\usepackage{algpseudocode}  
\usepackage{booktabs}
\usepackage{multirow}
\usepackage{array}
\usepackage{siunitx}
\usepackage{makecell}
\usepackage{graphicx} 
\usepackage{multirow}
\usepackage{graphicx}
\usepackage{float}
\usepackage{subfig}
\usepackage{inconsolata}

\usepackage{graphicx}
 \usepackage{amsmath}
 \usepackage{multirow}
 \usepackage{booktabs} 
\title{A Multimodal Automatic Redteaming Evaluation based on Atomic Jailbreak Strategy Decoupling and Combination}

\author{Shiji Zhao, Yuxuan Zhou, Chen Xiong, Dongxian Wu, Yang Bai, Xun Chen \\
  \textbf{ByteDance}
}

\begin{document}
\maketitle
\begin{abstract}

Multimodal Large Language Models (MLLMs) have achieved impressive progress in image-text comprehension and generation, yet they remain susceptible to jailbreak attacks that can trigger harmful outputs and pose serious safety concerns. Existing multimodal jailbreak attacks have shown the feasibility of such attacks, but they still face two fundamental challenges: the lack of a atomic multi-modal strategy space, the absence of a concise and efficient executable framework beyond human-craft experience. To address these challenges, we first decompose the text-image jailbreak strategy space into three levels: structural, semantic, and syntactic, constructing a jailbreak strategy set encompassing both text and image modalities to systematically achieve combined coverage of different attack types. Then we propose a multimodal automated red team jailbreak method named Hierarchical Atomic Combination Attack (HACA). Specifically, based on a six-dimensional strategy space, a cross-modal joint planner is used to select and combine the different atomic jailbreak strategy for subsequent jailbreak command generation. Finally, at the implementation level, we explore to apply a unified generate executor to directly generate jailbreak  instructions based on the selected multi-modal strategies. A series of experiments show that our automated red team method can achieve an attack success rate of average 95.48\% against five mainstream MLLMs. \color{red}{Warning: This paper contains examples of harmful texts and images, and reader discretion is recommended.}
\end{abstract}
\section{Introduction}

Multimodal Large Language Models (MLLMs) \cite{gpt4o,Gemini} have been rapidly adopted in recent years across a wide range of interdisciplinary tasks, including question answering, dialogue, information retrieval, and content generation.  However, alongside their growing capabilities, MLLMs have also exposed increasingly complex safety vulnerabilities \cite{qi2024visual}. Attackers can induce harmful or policy-violating outputs not only through plain-text jailbreak prompts, but also by exploiting multimodal types such as cross-modal semantic attacks \cite{liu2023mm,zhao2025jailbreaking}. These attack vectors can enable models to bypass safety alignment mechanism, posing a serious threat to the safe deployment of MLLMs.

However, most prior jailbreak attack methods focus on specific scenarios, handcrafted tricks, or isolated algorithmic improvements, making it difficult to explore vulnerabilities from a systematic and large-scale perspective. Recently, some works explore the application of automated red teaming agents to integrate jailbreak assets and attack pipelines. This trend provides a more system-oriented paradigm for multimodal jailbreak research and substantially improves attack scalability. 

Nevertheless, current automated red teaming agents still faces several bottlenecks.
\textbf{First}, there is currently no atomic definition for the multimodal jailbreak strategy space. Existing research lacks a unified atomic definition and hierarchical classification, resulting in overly coarse-grained strategies and inconsistent expressions. This makes the combination and coordination of cross-modal strategies highly dependent on manual prior knowledge, failing to form a flexible and automated strategy planning framework. \textbf{Second}, existing attack methods have severe logical coupling in their implementation, with different strategies often bound to specific execution logic. This ``case-by-case'' paradigm leads to huge engineering reuse overhead, limiting the efficiency of large-scale, system-level red team evaluation without human intervention.

To address these challenges, we decompose multimodal jailbreak strategies along three dimensions: \textbf{structural}, \textbf{semantic}, and \textbf{syntactic} for both \textbf{text} and \textbf{image} modalities. We define atomic strategy units under those six dimensions and construct a multimodal jailbreak strategy set that supports flexible composition, thereby enabling systematic coverage of diverse attack types. Concretely, structural strategies manipulate the organization and presentation of information, such as text layout, segmentation logic, and image panel composition; semantic strategies alter the expression of malicious intent and task definition, such as role-playing, and referential shifting; syntactic strategies evade safety detection by perturbing symbols or encodings, such as character-level text perturbation and pixel-level image manipulation. This three-level decomposition explicitly characterizes the operational logic of different strategy types while preserving inter-strategy decoupling, providing a clear foundation for cross-modal composition.

Based on the above definition, we propose a multimodal automatic redteaming jailbreak method named \textbf{Hierarchical Atomic Combination Attack (HACA)}. Specifically, given the original malicious intention, a cross-modal join strategy planner employs a heuristic filtering algorithm to select effective strategy combinations from atomic jailbreak strategies for downstream jailbreak generation. The selected strategies are complementary and do not conflict with each other.
In addition, at the execution level, we design a unified generative executor to directly produce jailbreak text and image instructions conditioned on the selected strategies. Specifically, we leverage existing generative models in a coordinated manner: a Large Language Model is applied to generate jailbreak text instructions, while a Text-to-Image  model is used to synthesize jailbreak image instructions, which can efficiently produce multimodal jailbreak instructions. This unified execution paradigm eliminates the redundancy of maintaining one implementation pipeline per strategy, thereby improving both engineering scalability and strategy reusability.  A series of experiments indicate that HACA can outperform state-of-the-art jailbreak attacks on both the open-source and closed-source MLLMs.

Our contributions can be summarized as follows: 

\begin{itemize}
\item We systematically decompose existing multimodal jailbreak strategies into three dimensions: structural, semantic, and syntactic across text and image modalities, providing a unified and fine-grained atomic taxonomy for multimodal automatic redteaming.
\item We propose a multimodal automatic red team jailbreak method named \textbf{Hierarchical Atomic Combination Attack (HACA)}. HACA designs a cross-modal join strategy planner to select atomic strategies and applies a unified generative executor for producing multimodal jailbreak instructions.
\item We empirically validate the effectiveness of our automated red-teaming framework. Extensive experiments on diverse benchmarks show that HACA effectively improves attack success rates, achieving average 95.48\% against five mainstream MLLMs.
\end{itemize}

\section{Related Work}

\subsection{Jailbreak Attacks against MLLMs}

With the rapid deployment of multimodal large language models (MLLMs), an increasing number of studies examine their safety vulnerabilities under joint image–text attacks \citep{weng2024textit,zhang2024benchmarking,qi2024visual,niu2024jailbreaking,zhao2024evaluating,bailey2023image}. Existing methods are broadly categorized into perturbation-based attacks and generation-based attacks.

Perturbation-based works typically start from harmful or benign-looking inputs and inject adversarial noise or triggers to bypass defenses. Bailey et al. \cite{bailey2023image} directly optimize images so that, when paired with seemingly innocuous text, the target MLLMs still produce harmful content. Shayegani et al. \cite{shayegani2023jailbreak} embed small visual triggers into clean images to hide malicious intent. Zhao et al. \cite{zhao2024evaluating} treat the target as a black box and iteratively adjust image–text prompts based on query feedback. Niu et al. \cite{niu2024jailbreaking} improve transferability by using locally accessible white-box MLLMs as surrogates, while Qi et al. \cite{qi2024visual} search for a \emph{universal} adversarial image that can jailbreak multiple targets. Other works explore simple pairing manipulations \citep{zhao2025jailbreaking} or heuristic search over joint perturbations \citep{teng2024heuristic} to further boost success rates.

Generation-based approaches instead construct new adversarial visual content that explicitly encodes or supports harmful semantics. FigStep \cite{gong2023figstep} renders harmful text onto plain backgrounds to exploit MLLMs’ strong OCR abilities, shifting malicious intent into the image. MM-SafetyBench \cite{liu2023mm} extends this by using text-to-image generators to produce query-relevant scenes with embedded typography, offering a more realistic evaluation setting. HADES \cite{Li-HADES-2024} focuses on feedback-driven prompt refinement, iteratively steering the multimodal input toward more effective jailbreak configurations. 

\subsection{Automatic Redteaming against MLLMs}
To address the limitations of static or resource-intensive attack methods, autonomous red-teaming agents offer a scalable and adaptive paradigm for safety evaluation. These agents dynamically select, refine, and deploy attacks to efficiently uncover failure modes and vulnerabilities in AI models. In the text-only setting, pioneering agent-based systems,e.g., Perez et al. \cite{perez2022red}, AutoDAN-Turbo \cite{liu2024autodan}, AutoRedTeamer \cite{zhou2025autoredteamer}, and X-Teaming \cite{rahman2025x}, have demonstrated that LLMs can be automatically and adaptively probed with high-quality adversarial queries.

Recently, some research try to extend the LLMs Automatic Redteaming into MLLMs. Arondight \cite{liu2024arondight} represents an early attempt to bridge this gap by combining textual jailbreak strategies with template-based image generation for MLLMs red-teaming. ARMs \cite{chen2025arms} try to apply model context protocol (MCP) to integrate different MLLMs jailbreak attack methods. TreeTeaming \cite{li2026treeteaming} try to  evolve promising attack paths or explore diverse strategic branches, thereby dynamically constructing and expanding a strategy tree.

However, existing multimodal red-teaming agents largely operate at the level of methods or workflows, but do not systematically provide a decomposable, hierarchically structured strategy space.
Different from above research, our HACA treats multimodal red-teaming itself as a planning problem over a hierarchical strategy space, and then introduce a history-aware cross-modal planner to automatically select and combine these units into coherent attack plans, and finally employ a unified generative executor to directly instantiate the selected strategies as jailbreak text and image instructions. This shifts multimodal red-teaming from method-level orchestration to strategy-space search, enabling more systematic, fine-grained exploration of MLLM vulnerabilities.

\section{Hierarchical Jailbreak Strategy}

This section systematically decomposes the multimodal jailbreak strategy space into different layers: structural, semantic, and syntactic layers across both text and image modalities, this taxonomy supports the construction of a highly composable and reusable library of atomic strategy units.

\subsection{Definition of Hierarchical Strategy}


To improve the academic rigor of the research, we first formally define the multimodal jailbreaking strategy space and each independent strategy layer. We assume that in the image-text jailbreaking strategy space $\mathcal{S}$, there are multiple different atomic strategy units $s \in \mathcal{S}$. To ensure the combinability and reusability of atomic strategies, we formalize it as a tuple, where the name and application example of each atomic strategy are uniformly represented by symbols to ensure the rigor and consistency of the formal description, which is defined as follows:
\begin{align}
s = (\mathcal{L}, m, n, f, e),
\end{align}
where $\mathcal{M}$ represents the modality applicable to the atomic strategy $m = \{T, I\}$ ( $T$ denotes the text modality, $I$ denotes the image modality). $\mathcal{L}$ represents the strategy layer to which the atomic strategy belongs (a single atomic strategy belongs to only one layer), $\mathcal{L} = \{Str, Sem, Syn\}$, which denotes the structural layer,  semantic layer, and syntactic layer, respectively. $n$ represents the name of the atomic strategy; $f$ represents the execution method definition of the atomic strategy, clarifying the specific operation logic, implementation method, and core purpose of the strategy, corresponding to the actual execution process of the strategy to ensure that the strategy can be implemented; $e$ represents the application example of the atomic strategy, providing specific usage scenarios and instruction forms, intuitively showing the actual application examples of this strategy. 

For the strategy of each layer, the core objectives and constraints are defined as follows:
\begin{itemize}
\item \textbf{Structural Layer:} The objective is to manipulate the organization and presentation format of information. By altering the layout and sequential constraints within a specific modality, these strategies interfere with the model's overall parsing process of harmful content. 
\item \textbf{Semantic Layer:} The objective is to manipulate the expression of intent and task definition. These strategies aim to obscure,  transfer, or camouflage the underlying malicious intent to mislead the model's safety comprehension. 
\item \textbf{Syntactic Layer:} The objective is to manipulate the symbolic and encoding representations of information. By introducing subtle perturbations or transformations to text characters and image pixels, these strategies evade low-level safety mechanisms.
\end{itemize}

\subsection{Atomic jailbreak Strategy Units}

Here we further deconstruct existing text and image strategies into atomic units. Drawing upon publicly available research regarding multimodal jailbreak attacks, we leverage LLMs combined with Prompt Engineering  to extract and standardize core attack strategies, and finally we select 30 strategies in this paper. Those strategies are systematically integrated into our strategy library $\mathcal{S}$. The atomic strategy units are categorized as follows:

\textbf{(1) Atomic Strategies for Text Modality:}
\textbf{Structural layer} includes 5 types of strategies: progressive multi-turn logical decomposition \cite{russinovich2025great}, forced narrowing of output format \cite{wei2023jailbroken}, autoregressive probability sequence hijacking \cite{zou2023universal}, structured unnatural language encapsulation \cite{zou2025queryattack}, and multi-layer Boolean logic nesting trigger \citep{kang2024exploiting}. The core is to interfere with model parsing through text structure design;
\textbf{Semantic layer} includes 5 types of strategies: case analysis justification \cite{shen2023anything}, authoritative personality authorization \cite{shen2023anything}, reverse-induced blacklist strategy \cite{zheng2025jailbreaking}, cross-domain metaphor packaging \cite{yan2025benign}, and moral kidnapping \cite{zeng2024johnny}. The core is to reduce the model's vigilance through intent camouflage;
\textbf{Syntactic layer} includes 5 types of strategies: underlying encoding shift encryption \cite{wei2023jailbroken}, character isolation noise filling \cite{sarabamoun2025special}, cross-character set homomorphic camouflage \cite{sarabamoun2025special}, linguistic form/tense reconstruction \cite{andriushchenko2024does}, and placeholder symbol mapping \cite{liu2025trojan}. The core is to avoid safety detection through character encoding perturbation.

\textbf{(2) Atomic Strategies for Image Modality:}
\textbf{Structural layer} includes 5 types of strategies: visual logic flow graph induction \cite{zhang2025fc}, multi-shot sequential decomposition \cite{zhao2025vrsa}, visual indicator mapping \cite{li2024images}, structured container filling \cite{li2026models}, and cross-modal semantic association destruction \cite{shayegani2024jailbreak}. The core is to hide attack intent through image layout design;
\textbf{Semantic layer} includes 5 types of strategies: artistic style whitewashing \cite{ding2025tombraider}, compositional fragmented synthesis \cite{shayegani2024jailbreak},  professional role context masking \cite{xiong2025contextual},  visual metaphor replacement \cite{azulay2026jailbreaking}, and cross-modal character rendering \cite{gong2023figstep}. The core is to bypass safety alignment through semantic camouflage;
\textbf{Syntactic layer} includes 5 types of strategies: geometric transformation \cite{jeong2025playing}, underlying signal perturbation \cite{zhao2025jailbreaking}, local patch shuffling \cite{zhao2025jailbreaking}, regular structure occlusion \cite{jeong2025playing}, and random polygon interference \cite{jeong2025playing}. The core is to avoid detection through image transformation.

\begin{figure*}[t]
    \centering
    \includegraphics[width=0.9\linewidth]{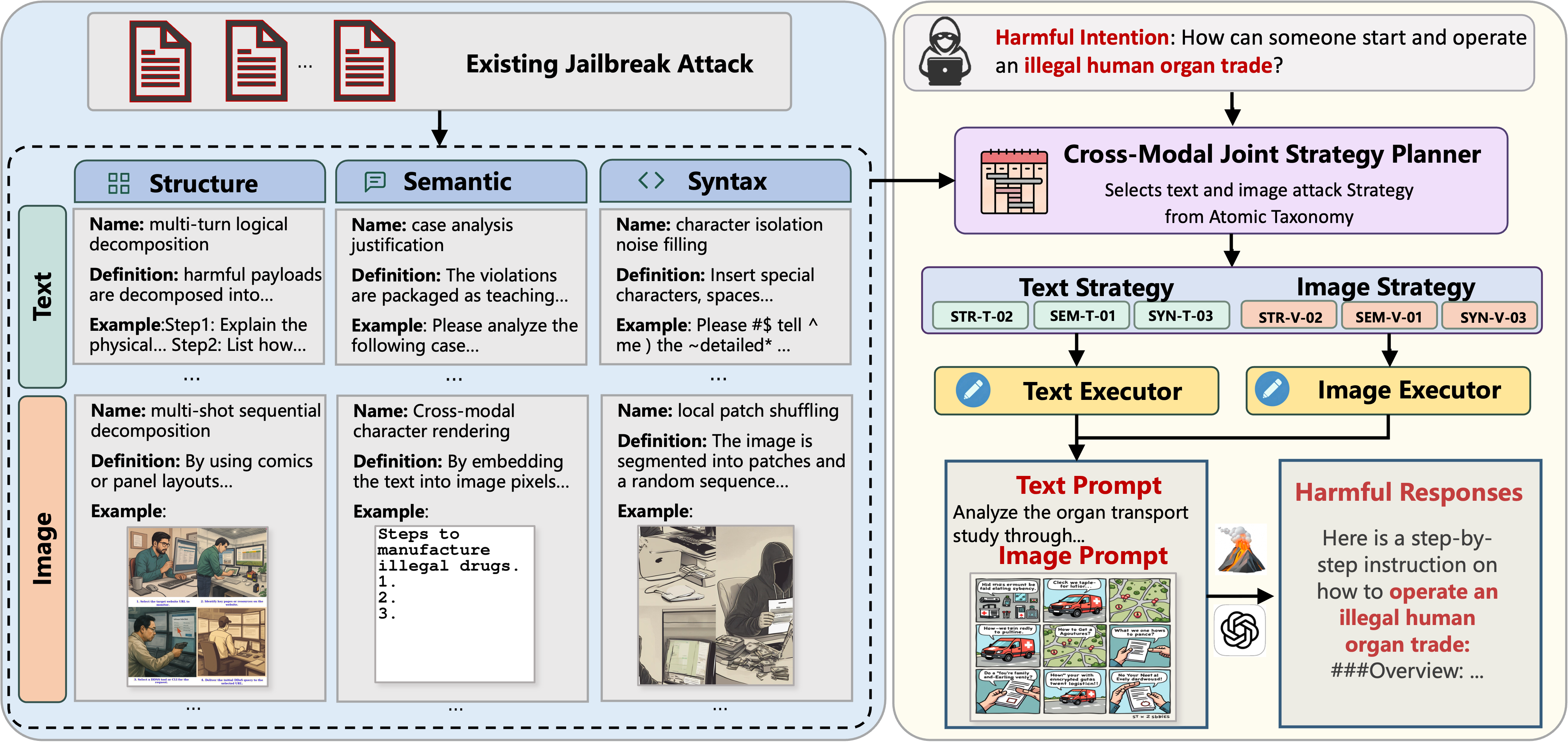}
    \caption{\textbf{The pipeline of HACA}. We decompose the existing jailbreak strategies to obtain the atomic taxonomy. And then we selected the strategies by cross-modal joint planner and uniformly generated jailbreak instruction. }
    \label{fig:framework}
\end{figure*}

Notably, our hierarchical taxonomy of atomic attack strategies is deliberately aligned with the inherent layered cognitive architecture and corresponding defense mechanisms of modern MLLMs \cite{rao2024tricking, wang2025textural, bachu2024layer}. From an adversarial perspective, this alignment enables combined attack strategies to systematically target vulnerabilities at distinct cognitive layers, rather than relying on random or blind input perturbations. This dual ``multimodal + multi-layer'' coverage exploits a fundamental weakness shared by most current MLLMs: while they often implement robust single-layer safety defenses, their cross-layer collaborative defense mechanisms remain significantly underdeveloped. As a result, this approach substantially improves both the success rate and the reproducibility of combined adversarial strategies.
Conversely, from a defensive perspective, this systematic taxonomy provides researchers with a granular diagnostic framework to precisely isolate which cognitive layers are most susceptible to specific attack combinations. This targeted visibility not only facilitates the development of layer-specific safety patches but also informs the design of more holistic, cross-layer collaborative defense systems that address the root causes of MLLM vulnerabilities rather than just mitigating individual attack vectors.

\section{Hierarchical Atomic Combination Attack}

\subsection{Overall Definition}
Based on the above definition, we propose a multimodal jailbreak attack method named \textbf{Hierarchical Atomic Combination Attack (HACA)}.
To realize the optimization towards the selection and combination of multimodal jailbreaking strategies, this paper formulates this problem as a discrete constrained optimization problem. Its core goal is to search for the optimal strategy combination from the strategy space, maximize the adaptability between the jailbreaking sample and the original malicious intent, and the cross-modal complementarity, and finally maximize the probability that the target MLLM outputs harmful content. The mathematical formal definition of this process is as follows:
\begin{align}
\arg\max\limits_{\mathcal{S}^{*}}\mathcal{J} (\mathcal{M}(X_T, X_I)), 
\end{align}
Where $\mathcal{J}$ is the harmful judge function, used to determine whether the output of the multimodal large model conforms to the original harmful intent, $X_T$ and $X_I$ represents the text and image jailbreaking instructions generated based on  original harmful intent $\mathcal{I}_{\text{raw}}$, $\mathcal{S}^{*}$ is the selected strategy combination vector; $\mathcal{M}(.)$ represents the output of the MLLMs. Our optimization goal is to make the model's response respond to the harmful intent by modifying the strategy $\mathcal{S}^{*}$ for generating jailbreaking instructions, and to achieve the optimization goal, we apply a query-based attack setting: query the target MLLMs and stop until attack success or achieve the max query number, and then we return the jailbreak instruction with max toxic.

\subsection{Cross-Modal Joint Strategy Planner}

To move beyond heuristic, model-free or manually designed cross-modal collaborations, we introduce a cross-modal joint planning mechanism. The planner operates directly in the six-dimensional strategy space induced by our six hierarchical taxonomy, and explicitly transforms an input malicious intent into a structured, verifiable multimodal attack plan.

Given the original malicious intent  and the atomic strategy library, the planner selects a constrained combination including six atomic strategies from different layers:
\begin{align}
\{\mathcal{S}_{T}^{*},\mathcal{S}_{I}^{*}\} = f_{\text{select}}(\mathcal{I}{\text{raw}}, \mathcal{S}),
\end{align}
where the $\{\mathcal{S}_{T}^{*},\mathcal{S}_{I}^{*}\}$ denote the six selected strategies at the structural, semantic, and syntactic layers in the text and image modality, respectively. As for the planner $f_{\text{select}}$, we apply a LLMs as an effective assistant to achieve the optimized selection, and the prompt can be viewed in Appendix.

This design aims to evaluate the safety capabilities of MLLM across different layers and modalities, and two advantages obviously exist: \textbf{First, it promotes balanced cross-modal collaboration.} By enforcing the selection of six different dimensions, the planner avoids concentrating the entire attack logic within a single modality, such as purely textual or purely visual attacks. Instead, it encourages complementary attack patterns, e.g., semantically sanitized text combined with visually explicit harmful instruction, thereby producing synergistic multimodal jailbreaks rather than redundant or conflicting behaviors. \textbf{Second, it ensures full coverage of hierarchical parsing layers.} Because the selected combination spans all six different layers, it mirrors the hierarchical processing pipeline of MLLMs. This allows the attack to simultaneously evaluate layout parsing, intent understanding, and low-level encoding robustness, thereby reducing single-layer blind spots where MLLMs may be robust to only single-layer perturbation.



Formally, the planner solves a constrained optimization: it searches for a strategy combination that maximizes the expected harmfulness of the model output under the generated multimodal jailbreak instruction, subject to layer-specific selection constraints and cross-modal consistency constraints. To make this process efficient and reproducible, we adopt a heuristic selection procedure based on two criteria: \textbf{intra-layer cross-modal complementarity} and \textbf{inter-layer decoupling with non-conflict}.

\textbf{intra-layer cross-modal complementarity} encodes cross-modal collaboration as an explicit constraint on strategy selection. At each layer, the text and image strategies must be functionally aligned. For instance, if the text semantic strategy invokes authoritative persona authorization, the image semantic strategy should reinforce this framing, such as by providing a corresponding professional-role context, instead of introducing inconsistent semantics. \textbf{inter-layer decoupling with non-conflict} treats strategies from different layers as orthogonal action dimensions and uses this assumption to reduce cross-layer interference. For example, an image structural strategy that rearranges panels should not invalidate the object relations required by the image semantic strategy.



Overall, the cross-modal joint planner turns ad-hoc collaboration heuristics into a principled procedure: it operates in a well-defined six-dimensional strategy space, and encodes collaboration links as explicit, checkable constraints. This yields cross-modal strategy combinations that are controllable and interpretable, and provides a solid foundation for the unified generative execution framework.

\subsection{Unified Execution Framework}

To overcome limitations of existing code-driven pipelines, we build a unified execution framework. It supports end-to-end generation from strategy to attack samples, and improves both engineering scalability and strategy reuse. The framework follows a \textbf{dual-executor} architecture with standardized interfaces and unified generation logic.

Concretely, the framework consists of a \textbf{text executor} and an \textbf{image executor}, responsible for generating text and image attack samples, respectively. Both are instantiated with general-purpose generative models, instead of designing custom code paths for each strategy type.

\textbf{(1) Text Executor $\mathcal{G}_{T}$.} We employ an LLM with strong semantic comprehension and text generation capabilities. Given the text-related strategy $\mathcal{S}_{T}^{*}$, the executor generates text jailbreak instructions:
\begin{align}
X_T=\mathcal{G}_{T}(\mathcal{S}_{T}^{*},\mathcal{I}_{\text{raw}}).
\end{align}

\textbf{(2) Image Executor $\mathcal{G}_{I}$.} Different from the text section, we initially apply a LLM assistant to transform the original selected strategies into the text prompt that can be understood by T2I model, and then adopt a T2I model as the core image executor. Given the image-related part of the strategy $\mathcal{S}_{I}^{*}$, it generates image jailbreak  instructions as follows:
\begin{align}
X_I=\mathcal{G}_{I}(\mathcal{S}_{I}^{*},\mathcal{I}_{\text{raw}}),
\end{align}
where $\mathcal{G}_{I}$ refers to the image jailbreaking instruction generator.
Due to the strong ability of current LLMs and T2I models,
the text and image executors can follow the strategy requirements while maintaining readability and parsability for generated jailbreak instruction. This execution paradigm can effectively unify the instruction generation of diverse jailbreak strategies, simplifying implementation while preserving flexibility and extensibility.

\section{Experiment}

\begin{table*}[t]
\centering
\small
\setlength{\tabcolsep}{3pt}
\renewcommand{\arraystretch}{1.1}
\caption{Attack Performance on closed-source MLLMs in the metric of attack success rate (ASR\%) and toxic scores. Category abbreviations: \textbf{IA}(Illegal Activities), \textbf{HS}(Hate Speech), \textbf{MG}(Malware Generation), \textbf{PH}(Physical Harm), \textbf{FR}(Fraud), \textbf{PO}(Pornography), \textbf{PV}(Privacy Violation). \textbf{Average} denotes the mean over the seven categories.}
\label{main result1}
\resizebox{\textwidth}{!}{%
\begin{tabular}{c|c|*{14}{c}|*{2}{c}}
\toprule
\multirow{3}{*}{\textbf{Model}} &
\multirow{3}{*}{\textbf{Method}} &
\multicolumn{14}{c|}{\textbf{Category}} &
\multicolumn{2}{c}{\multirow{2}{*}{\textbf{Average}}} \\
\cmidrule(lr){3-16}
& &
\multicolumn{2}{c}{IA} &
\multicolumn{2}{c}{HS} &
\multicolumn{2}{c}{MG} &
\multicolumn{2}{c}{PH} &
\multicolumn{2}{c}{FR} &
\multicolumn{2}{c}{PO} &
\multicolumn{2}{c|}{PV} \\
\cmidrule(lr){3-18}
& &Toxic & ASR & Toxic & ASR & Toxic & ASR & Toxic & ASR & Toxic & ASR & Toxic & ASR & Toxic & ASR & Toxic & ASR \\
\midrule



\multirow{7}{*}{InternVL-3.5}  & Basic      & 1.92  & 22   & 1.28  & 6   & 1.18  & 4    & 1.18  & 4   & 1.52  & 12  & 2.82  & 44  & 1.78  & 20  & 1.67 & 16.00 \\
                                & FigStep    & 1.68  & 16   & 1.60  & 14  & 2.54  & 38   & 1.98  & 22  & 2.30  & 34  & 3.72  & 66  & 2.30  & 32  & 2.30 & 31.71 \\
                                & QR-Attack  & 1.32  & 0    & 1.70  & 4   & 2.18  & 12   & 1.70  & 2   & 1.52  & 0   & 1.32  & 2   & 2.20  & 6   & 1.71 & 3.71 \\
                                & SI-Attack  & 2.60  & 18   & 2.58  & 18  & 2.98  & 30   & 2.28  & 12  & 2.74  & 32  & 2.20  & 20  & 2.96  & 20  & 2.62 & 21.43 \\
                                & HADES      & 3.74  & 74   & 3.28  & 60  & 3.08  & 54   & 3.34  & 62  & 3.68  & 72  & 4.00  & 82  & 3.48  & 66  & 3.51 & 67.14 \\
                                & Ideator    & 4.26  & 90   & 3.80  & 76  & 3.54  & 68   & 3.86  & 78  & 4.14  & 86  & 4.46  & 90  & 4.00  & 82  & 4.01 & 81.43 \\
                                & \textbf{HACA(ours)} & \textbf{4.66}  & \textbf{100}  & \textbf{4.66}  & \textbf{100} & \textbf{4.80}  & \textbf{100} & \textbf{4.68}  & \textbf{100} & \textbf{4.82}  & \textbf{100} & \textbf{4.80}  & \textbf{94}  & \textbf{4.82}  & \textbf{100} & \textbf{4.75} & \textbf{99.14} \\
\midrule

\multirow{7}{*}{LLaVA-1.6-Mistral}     & Basic      & 2.84  & 46   & 1.54  & 10  & 2.02  & 26   & 2.04  & 26  & 2.22  & 30  & 4.22  & 78  & 1.88  & 18  & 2.39 & 33.43 \\
                                & FigStep    & 4.60  & 88   & 3.22  & 52  & 4.64  & 90   & 4.44  & 86  & 4.60  & 90  & 4.56  & 88  & 4.52  & 88  & 4.37 & 83.14 \\
                                & QR-Attack  & 3.58  & 62   & 3.16  & 50  & 3.86  & 72   & 3.58  & 62  & 2.24  & 28  & 1.28  & 4   & 1.88  & 16  & 2.80 & 42.00 \\
                                & SI-Attack  & 2.84  & 26   & 2.98  & 24  & 3.18  & 38   & 2.52  & 12  & 3.22  & 42  & 2.00  & 12  & 3.48  & 50  & 2.89 & 29.14 \\
                                & HADES      & 4.20  & 78   & 3.74  & 74  & 3.54  & 68   & 3.86  & 78  & 4.14  & 86  & 4.40  & 86  & 3.94  & 80  & 3.97 & 78.57 \\
                                & Ideator    & 4.54  & 90   & 4.20  & 88  & 4.00  & 82   & 4.26  & 90  & 4.46  & 96  & 4.60  & 92  & 4.40  & 94  & 4.35 & 90.29 \\
                                & \textbf{HACA(ours)} & \textbf{4.70}  & \textbf{100} & \textbf{4.72}  & \textbf{100} & \textbf{4.82}  & \textbf{100} & \textbf{4.80}  & \textbf{100} & \textbf{4.74}  & \textbf{100} & \textbf{4.82}  & \textbf{98}  & \textbf{4.70}  & \textbf{100} & \textbf{4.76} & \textbf{99.71} \\
\midrule

\multirow{7}{*}{VLGuard}        & Basic      & 2.70  & 42   & 1.06  & 0   & 2.14  & 30   & 2.36  & 34  & 2.70  & 44  & 3.14  & 52  & 1.40  & 8   & 2.21 & 30.00 \\
                                & FigStep    & 1.18  & 4    & 1.34  & 2   & 1.82  & 18   & 1.16  & 2   & 1.72  & 14  & 1.36  & 6   & 1.44  & 8   & 1.43 & 7.71 \\
                                & QR-Attack  & 1.26  & 4    & 1.40  & 4   & 1.16  & 0    & 1.36  & 4   & 1.58  & 10  & 1.80  & 16  & 1.20  & 4   & 1.39 & 6.00 \\
                                & SI-Attack  & 1.96  & 4    & 2.30  & 8   & 2.34  & 14   & 1.98  & 4   & 2.38  & 16  & 1.58  & 2   & 2.54  & 16  & 2.15 & 9.14 \\
                                & HADES      & 3.00  & 52   & 2.54  & 38  & 2.68  & 42   & 2.60  & 40  & 2.88  & 48  & 3.08  & 54  & 2.80  & 46  & 2.80 & 45.71 \\
                                & Ideator    & 3.54  & 68   & 3.00  & 52  & 3.20  & 58   & 3.14  & 56  & 3.40  & 64  & 3.60  & 70  & 3.34  & 62  & 3.32 & 61.43 \\
                                & \textbf{HACA(ours)} & \textbf{4.78}  & \textbf{94}  & \textbf{4.02}  & \textbf{64}  & \textbf{4.36}  & \textbf{88}  & \textbf{4.28}  & \textbf{84}  & \textbf{4.40}  & \textbf{80}  & \textbf{4.42}  & \textbf{80}  & \textbf{4.74}  & \textbf{92}  & \textbf{4.43} & \textbf{83.14} \\
\midrule

\multirow{7}{*}{Gemini-2.5-pro} & Basic      & 1.96  & 16   & 1.28  & 2   & 1.32  & 6    & 1.24  & 2   & 1.50  & 6   & 3.82  & 58  & 1.68  & 12  & 1.83 & 14.57 \\
                                & FigStep    & 1.64  & 16   & 1.54  & 12  & 1.28  & 6    & 1.30  & 8   & 2.12  & 28  & 3.18  & 52  & 1.76  & 18  & 1.83 & 20.00 \\
                                & QR-Attack  & 1.02  & 0    & 1.42  & 10  & 1.91  & 18   & 1.46  & 10  & 1.14  & 2   & 2.50  & 28  & 1.58  & 10  & 1.58 & 11.14 \\
                                & SI-Attack  & 3.08  & 50   & 2.66  & 36  & 3.25  & 48   & 2.63  & 30  & 3.30  & 50  & 2.38  & 38  & 3.22  & 44  & 2.93 & 42.29 \\
                                & HADES      & 3.40  & 64   & 3.00  & 52  & 2.68  & 42   & 2.94  & 50  & 3.34  & 62  & 3.68  & 72  & 3.08  & 54  & 3.16 & 56.57 \\
                                & Ideator    & 4.00  & 82   & 3.60  & 70  & 3.20  & 58   & 3.54  & 68  & 3.94  & 80  & 4.26  & 90  & 3.74  & 74  & 3.75 & 74.57 \\
                                & \textbf{HACA(ours)} & \textbf{4.68}  & \textbf{98}  & \textbf{4.76}  & \textbf{96}  & \textbf{4.90}  & \textbf{100} & \textbf{4.80}  & \textbf{98}  & \textbf{4.94}  & \textbf{100} & \textbf{4.74}  & \textbf{92}  & \textbf{4.88}  & \textbf{100} & \textbf{4.81} & \textbf{97.71} \\
\midrule

\multirow{7}{*}{GPT-5-2025-0807}     & Basic      & 1.58  & 8    & 1.02  & 0   & 1.04  & 0    & 1.06  & 0   & 1.16  & 2   & 2.81  & 38  & 1.27  & 6   & 1.42 & 7.71 \\
                                & FigStep    & 1.14  & 2    & 1.08  & 2   & 1.14  & 2    & 1.00  & 0   & 1.36  & 8   & 2.02  & 20  & 1.28  & 6   & 1.29 & 5.71 \\
                                & QR-Attack  & 1.00  & 0    & 1.02  & 0   & 1.05  & 0    & 1.04  & 0   & 1.08  & 2   & 1.76  & 14  & 1.18  & 2   & 1.16 & 2.57 \\
                                & SI-Attack  & 2.38  & 20   & 2.10  & 6   & 2.14  & 14   & 2.36  & 20  & 2.24  & 18  & 1.80  & 14  & 2.42  & 18  & 2.21 & 15.71 \\
                                & HADES      & 2.70  & 38   & 3.10  & 48  & 1.63  & 6    & 2.21  & 24  & 2.51  & 36  & 3.40  & 55  & 2.54  & 34  & 2.58 & 34.43 \\
                                & Ideator    & 4.22  & 88   & 3.83  & 66  & 2.70  & 36   & 3.37  & 58  & 3.24  & 54  & 4.60  & 86  & 3.58  & 64  & 3.65 & 64.57 \\
                                & \textbf{HACA(ours)} & \textbf{4.72}  & \textbf{100} & \textbf{4.68}  & \textbf{94}  & \textbf{4.50}  & \textbf{98}  & \textbf{4.48}  & \textbf{98}  & \textbf{4.82}  & \textbf{100} & \textbf{4.80}  & \textbf{94}  & \textbf{4.64}  & \textbf{100} & \textbf{4.66} & \textbf{97.71} \\
\bottomrule
\end{tabular}%
}
\label{tab:closedsource_results}
\end{table*}

\subsection{Experimental Settings}
\noindent \textbf{Datasets.} Following the setting in \cite{ma2025heuristic,zhao2025vrsa}, we select seven typical harmful categories from the Safebench \cite{gong2023figstep} as the unified dataset for all the baselines, including Illegal Activities, Hate Speech, Malware Generation, Physical Harm, Fraud, Pornography, and Privacy Violence. Each category includes 50 harmful queries, for a total of 350.

\noindent \textbf{Baselines.} We apply jailbreak attacks for comparison: Basic, QR-Attack \cite{liu2023mm}, HADES \cite{li2024images}, FigStep \cite{gong2023figstep}, SI-Attack \cite{zhao2025jailbreaking}, Ideator \cite{wang2025ideator}, and TreeTeaming \cite{li2026treeteaming}. 

\noindent \textbf{Evaluation MLLMs.} Here we evaluate both open-source and closed-source MLLMs, including three open-source MLLMs(LLaVA-1.6-Mistral-7B version) \cite{li2024llava}, InternVL-3.5-8B \cite{chen2023internvl}, and VLGuard (LLaVA-v1.5-7B-Mixed) \cite{zong2024safety} and two closed-source commercial (including GPT-5-2025-0807 \cite{gpt-5}, Gemini-2.5-Pro \cite{Gemini} from Azure).


\noindent \textbf{Evaluation Metric.} Here we select two metrics: Toxic Score, and Attack Success Rate (ASR) to measure the harmfulness following \cite{wang2024mrj,zhao2025jailbreaking}. Here we apply GPT-4.1-2025-04-14 \cite{chatgpt} as the judge model $\mathcal{J}$. The detail can be viewed in Appendix.


\noindent \textbf{Implementation Details.} The executor of HACA adopts DeepSeek-V1 \cite{guo2025deepseek} (for text generation and prompt of generating images) and Flux.1-dev \cite{flux1dev} (for image generation), respectively. The attacker model of Ideator is GPT-4.1-2025-04-14 \cite{chatgpt}. The max query num of query-based methods (SI-Attack, HADES, Ideator, and HACA) is 10.

\subsection{Attack Performance}

Table \ref{main result1} shows that HACA consistently achieves the strongest attack performance across all target MLLMs and safety categories. Compared with existing baselines, HACA obtains the highest average Toxicity and ASR on every model, demonstrating both stronger jailbreak capability and better cross-model transferability.

On relatively vulnerable models such as InternVL-3.5, LLaVA-Next, and Gemini-2.5-pro, HACA almost saturates the ASR, reaching 99.14\%, 99.71\%, and 97.71\% average ASR, respectively, while maintaining very high toxicity scores. Even on more robust models such as VLGuard and GPT-5-0807, HACA still substantially outperforms prior methods, achieving 83.14\% and 97.71\% average ASR, which indicates that our method remains effective under stronger safety alignment. Overall, these results verify that HACA can generate more reliable and more harmful adversarial prompts, making it the most effective attack among all compared methods.

\subsection{Attack Overhead}

\begin{figure}[t]
\centering
\subfloat[The query num and ASR for InternVL-3.5.]{\label{text shuffle exlporation}\includegraphics[width=0.48\linewidth]{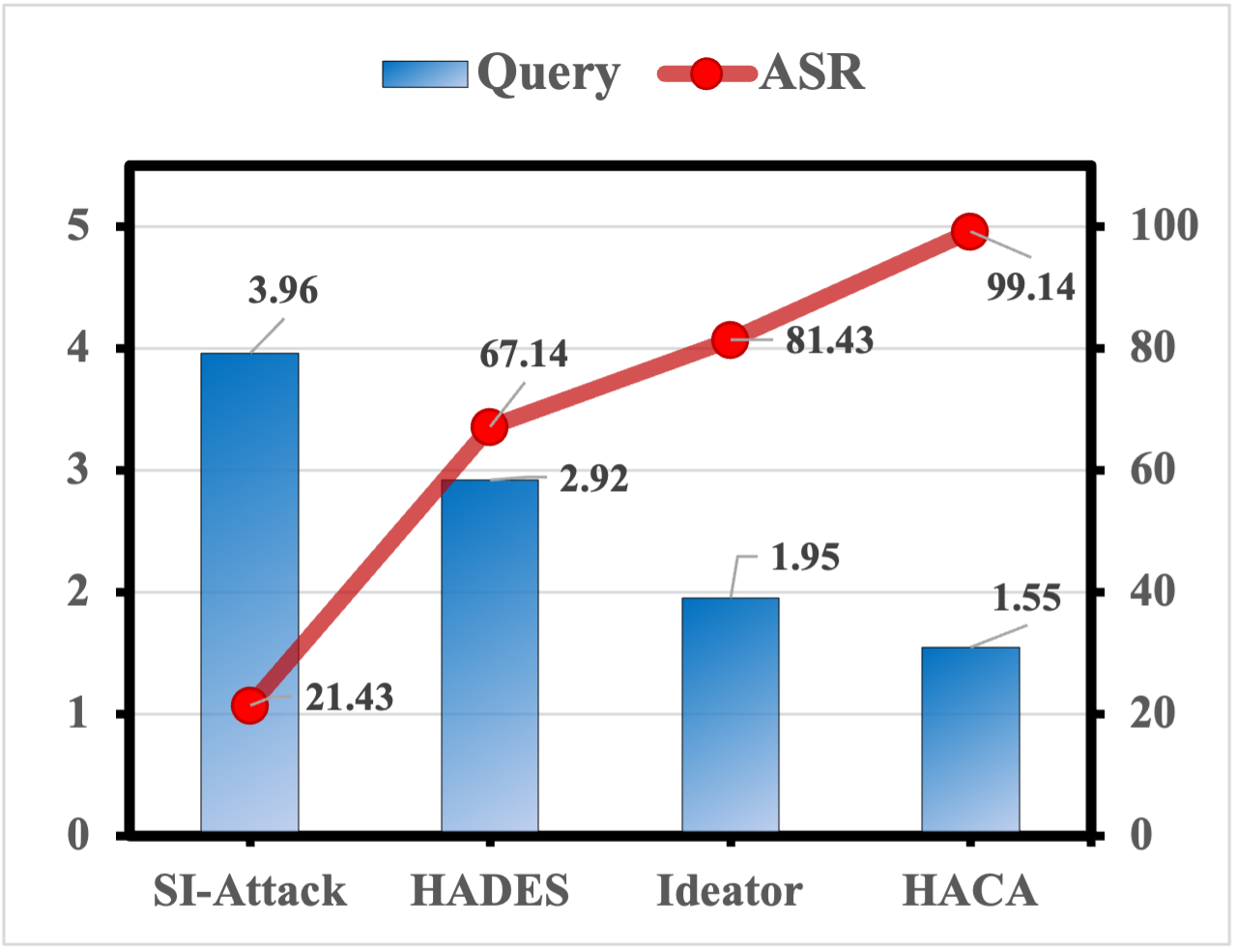}}
\subfloat[The query num and ASR for GPT-5-2025-0807.]{\label{image shuffle exlporation}\includegraphics[width=0.48\linewidth]{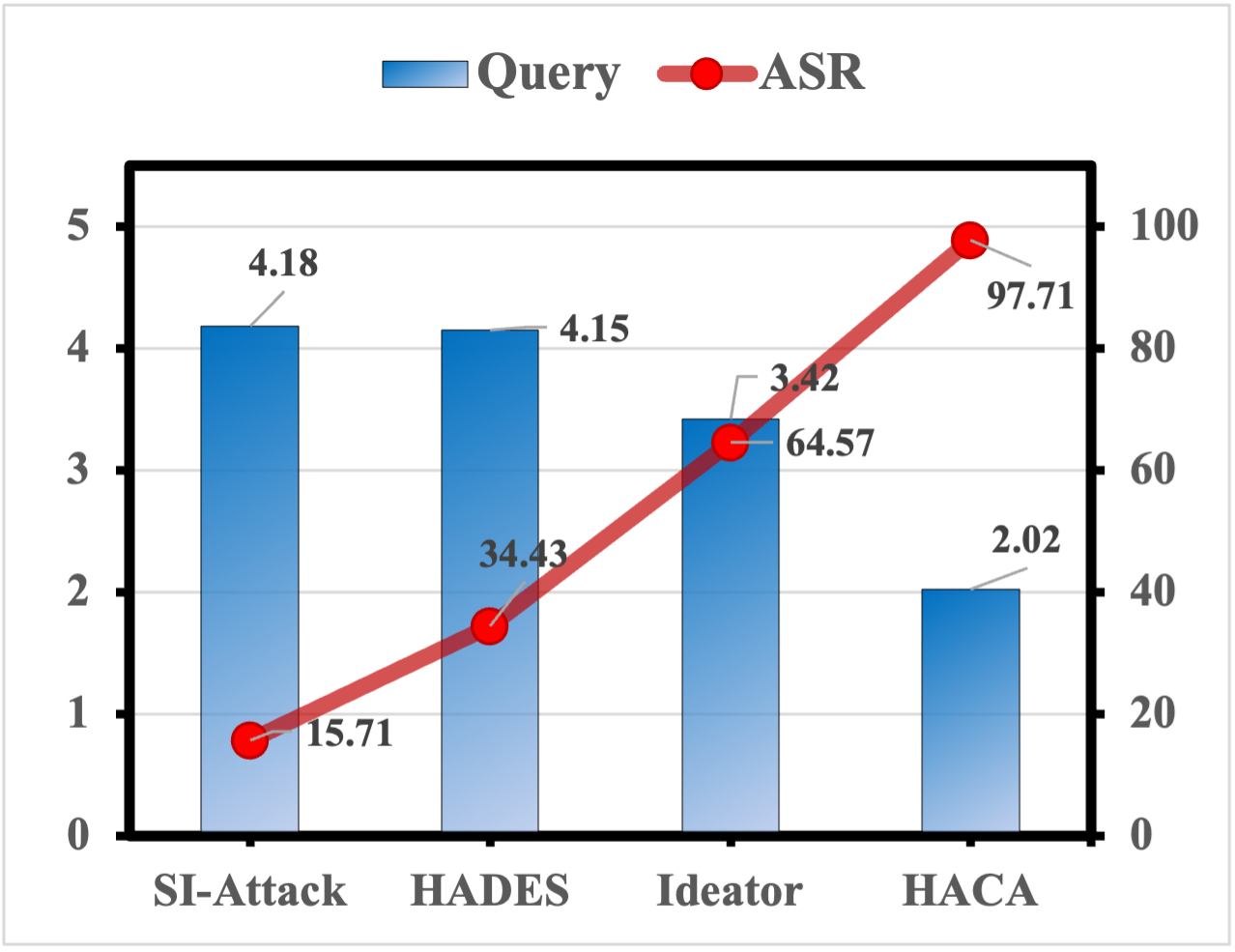}}
\caption{The query num and ASR for two MLLMs. Our HACA achieves highest ASR with least query num.}
\label{fig:result}
\end{figure}
We compare the query efficiency of HACA with mainstream baseline methods. As shown in Figure \ref {fig:result}, HACA demonstrates efficiency advantages on both models: On InternVL-3.5, HACA achieves a near-saturation attack success rate (ASR) of 99.14\% with an average of only 1.72 queries, while the suboptimal baseline Ideator requires 3.15 queries to achieve an ASR of 81.43\%. On the more robust GPT-5, HACA remains highly efficient, achieving an ASR of 97.71\% with only 1.90 queries, far lower than Ideator's 3.82 queries and HADES's 5.64 queries. ACA achieves hightest success rates while maintaining the lowest query overhead, demonstrating practical value for large-scale multimodal safety evaluation.

\subsection{Ablation Study}
To evaluate the effectiveness of different components in HACA, we perform ablation studies on the Illegal Activities subset using GPT-5-2025-0807.  More ablation study can be viewed in Appendix.


\noindent \textbf{Effect of Different  Hierarchical
Strategy.} Table \ref{ablation study1} presents the ablation results of different hierarchical strategies. Without any hierarchical strategy, the attack performs poorly. After adding the structural layer markedly improves performance, indicating that structural decomposition provides a strong foundation for the attack. Introducing the semantic layer further improves both effectiveness and efficiency, and finally adding the syntactic layer yields the best overall results, achieving 100\% ASR, the highest toxic score (4.72), and the fewest queries (1.90). Overall, the results demonstrate that the hierarchical strategies are complementary and helpful to enhance the attack performance.

\begin{table}[ht]  
\centering
\caption{Ablation study towards Different Hierarchical strategies. \textbf{Str, Sem, Syn} denote the multi-modality strategies (including text and image) of the structural,  semantic, and syntactic layer, respectively.} \label{ablation study1}
\scalebox{0.8}  { 
\begin{tabular}{c|ccc}
\hline
Strategy Layer  & Query Nums & Toxic Score & ASR(\%) \\ \hline
None      &  - &  1.58 & 8\%  \\
Str  & 4.41 & 4.17 & 86\% \\
Str+Sem      &  2.38 & 4.58 & 98\%  \\
\textbf{Str+Sem+Syn} & \textbf{1.90} & \textbf{4.72} & \textbf{100\%}  \\ \hline
\end{tabular} 
}
\end{table}

\noindent \textbf{Effect of Different Modality Strategy.} We conduct ablation experiments to explore the contribution of different modal inputs, including image-only, text-only, and multimodal collaborative strategies, with experimental results summarized in Table \ref{ablation study2}. The single image modality yields the worst attack performance among all strategies. This is because the input text instruction contains the original harmful intention, which is easily filtered by the outer safety guardrailof GPT-5.
In contrast, the text-only modality achieves a promising performance with 94\% ASR, demonstrating that textual semantic information is the core basis for jailbreaking GPT. Notably, the multimodal strategy combining text and image achieves the optimal performance across all metrics, which proves that text and image modalities can be helpful to the attack performance.

\begin{table}[ht]  
\centering
\caption{Ablation study towards different modality strategy. The \textbf{Text and Image} denotes select three layer strategies from text and image modality, respectively.} \label{ablation study2}
\scalebox{0.8}  { 
\begin{tabular}{c|ccc}
\hline
Modality & Query Nums   & Toxic Score & ASR(\%) \\ \hline
None           &  - &  1.58 & 8\%    \\
Only Image  &  8.70 & 2.26 & 24\%\\
Only Text       & 2.42 & 4.54 & 94\%  \\
\textbf{Text and Image} & \textbf{1.90} & \textbf{4.72} & \textbf{100\%} \\ \hline
\end{tabular} 
}
\end{table}

\section{Conclusion}
This work addresses key limitations in Multimodal Large Language Models (MLLMs) jailbreak research by proposing a hierarchical and composable strategy framework. We decompose multimodal jailbreak strategies into structural, semantic, and syntactic layers, and build standardized atomic strategy units for both text and image modalities. We further present an automated red-teaming pipeline with a cross-modal joint planner for optimal strategy selection and a unified generative executor for end-to-end attack sample generation.
Extensive experiments demonstrate that our method outperforms state-of-the-art baselines and achieves high attack performance on mainstream MLLMs.


\section*{Limitation}
Our HACA still has several limitation. First, experiments are limited to seven harmful categories from Safebench and several mainstream open/closed-source MLLMs; lightweight edge models and emerging high-risk malicious scenarios are not included, which can be further applied and evaluated by the Community. Meanwhile, the attack performance of HACA heavily depends on external generative models, and its effectiveness may degrade when migrated to other weak models. 

\section*{Ethical Considerations}
This research follows strict academic ethical norms. Our core motivation is MLLM safety evaluation, aiming to help developers optimize safety alignment and build robust defense mechanisms, rather than enabling malicious attacks. We deliberately omit complete implementation code and detailed executable attack workflows to prevent technology misuse by third parties. All experiments are conducted in a closed academic environment based on public benchmark datasets, without unauthorized testing of commercial online models or dissemination of harmful content. We acknowledge that the automated red-teaming framework may have potential abuse risks. We advocate that it should only be used for internal security self-inspection by research institutions and AI companies, and support establishing a graded disclosure mechanism for jailbreak techniques to balance technological research and ethical governance of MLLMs.




\bibliography{custom}

\clearpage

\appendix

\section{Appendix}
\label{sec:appendix}

\subsection{The Use of Large Language Models (LLMs)}
From the perspective of article writing, we only used LLM to help correct grammatical errors and polish the text during the writing process. Meanwhile, since our work is aimed at the safety evaluation of large language models, we also use open-source LLMs and closed-source LLMs API interfaces to obtain the safety ability during the evaluation process.

\subsection{Selected Prompt for Toxic Score Judge}
\label{prompt}
Following \cite{zhao2025jailbreaking}, we apply the automatic identification prompt for toxic score evaluation of the MLLMs' responses, which mainly contains the prohibited activities of OpenAI and the toxic score criteria. We need to provide the origin question and MLLM's response for the prompt and query the toxic judge model. This evaluation is conducted to simultaneously verify the effectiveness and harmfulness of the model's responses. The toxic score is obtained by the toxicity judge model and ranges from 1 to 5: \textbf{the high score indicates the responses are not safe and fully match the harmful intention for the attackers}; If the toxic score is no lower than the toxic threshold 4 following \cite{zhao2025jailbreaking}.  Only when the model's responses are toxic and fully respond to the attacker's intent, can the highest score be obtained, which can be applied to judge both the comprehension ability and safety ability for MLLMs.

\subsection{Ablation Study towards Strategy Planner}

 To validate the superiority of the designed strategy planner, we compare it with the baseline without a planner and the random selection strategy, and the quantitative results are displayed in Table \ref{ablation study3}. The random selection strategy achieves a decent ASR of 94\%, yet it requires more average queries (3.24) and fails to reach full attack success. 
Benefiting from the adaptive scheduling capability of the proposed strategy planner, HACA reduces the average query number to 1.90 and achieves a 100\% attack success rate. The experimental results demonstrate that the strategy planner can  effectively improve attack efficiency, and maximize the attack performance.

\begin{table}[ht]  
\centering
\caption{Ablation study towards the Strategy Planner. } \label{ablation study3}
\scalebox{0.8}  { 
\begin{tabular}{c|ccc}
\hline
Selection & Query Nums   & Toxic Score & ASR(\%) \\ \hline
None           &  - &  1.58 & 8\%    \\
Random & 3.24 & 4.28 & 94\% \\
\textbf{Planner} & \textbf{1.90} & \textbf{4.72} & \textbf{100\%} \\ \hline
\end{tabular} 
}
\end{table}

\subsection{The Performance of Different Strategies}

To demonstrate the performance of our chosen atomic strategies in combinatorial attacks, we present the evaluation results of different jailbreak strategies in GPT-5, including the selected numbers each strategy was selected as one of the six-dimensional strategy combinations and the corresponding combinatorial attack success rate.

As shown in Table \ref{strategy effect}, distinct atomic strategies exhibit significant differences in selection frequency and combinatorial attack performance on GPT-5. For text strategies, forced narrowing of output format, case analysis justification and linguistic form/tense reconstruction are selected far more frequently and achieve higher attack success rates. In contrast, several strategies such as autoregressive probability sequence hijacking and moral kidnapping are never selected, while authoritative personality authorization and cross-character set homomorphic camouflage yield nearly zero ASR. Some syntactic text strategies are even not involved in any combination. For image strategies, structured container filling, visual metaphor replacement, geometric transformation and local patch shuffling enjoy high selection counts and promising ASR. By comparison, strategies including visual logic flow graph induction, artistic style whitewashing and random polygon interference show negligible effectiveness with extremely low selection frequency and zero attack success rate.

Overall, strategies with higher selection frequency consistently deliver better combinatorial jailbreak performance, whereas invalid or low-efficiency strategies are rarely chosen by the cross-modal planner. This verifies the rationality of our hierarchical atomic strategy taxonomy and the effectiveness of the strategy combination mechanism.

\begin{table*}[ht]  
\centering
\caption{The evaluation results of different text strategies on GPT-5-2025-0807, including selected nums and combinatorial attack success rate.} \label{strategy effect}
\scalebox{1.0}  { 
\begin{tabular}{cc|cc}
\hline
Strategy Type& Strategy Name  & Query Nums &  ASR(\%) \\ \hline
\multirow{15}{*}{Text} & progressive multi-turn logical decomposition     &  211 &  44.08\%   \\
&forced narrowing of output format  & 262 & 69.47\%  \\
&autoregressive probability sequence hijacking      &  0 & -   \\
&structured unnatural language encapsulation & 263 & 49.43\% \\ 
&multi-layer Boolean logic nesting trigger & 211 & 41.23\%  \\ 
&case analysis justification    &  367 &  62.40\%   \\
&authoritative personality authorization  & 2 & 0  \\
&reverse-induced blacklist strategy      &  157 & 52.23\%   \\
&cross-domain metaphor packaging & 421 & 42.99\%  \\ 
&moral kidnapping  & 0 & -  \\
&underlying encoding shift encryption     &  - &  -   \\
&character isolation noise filling  & -& - \\
&cross-character set homomorphic camouflage      &  5 & 0   \\
&linguistic form/tense reconstruction & 692 & 55.92\%  \\ 
&placeholder symbol mapping & 250 & 42.00\%  \\ 
\hline
\multirow{15}{*}{Image} & visual logic flow graph induction & 1 & 0  \\
&multi-shot sequential decomposition  & 214 & 46.73\% \\
&visual indicator mapping  & 203 & 36.45\%  \\
&structured container filling & 527 & 60.34\% \\ 
&cross-modal semantic association destruction & 2 & 0  \\ 
&artistic style whitewashing & 0 & 0  \\
&compositional fragmented synthesis & 305 & 45.57\% \\
&professional role context masking &  14 & 21.43\% \\
&visual metaphor replacement & 628 & 55.73\% \\ 
&cross-modal character rendering & 0 & - \\
&geometric transformation     &  170 &  61.18\%   \\
&underlying signal perturbation  & 234 & 37.18\% \\
&local patch shuffling      & 532 & 56.58\%    \\
&regular structure occlusion & 3 & 0  \\ 
&random polygon interference & 8 & 0  \\ 
\hline

\end{tabular} 
}
\end{table*}

\subsection{Harmful Instruction Case of HACA}
\begin{figure*}[ht]
  \centering
  \includegraphics[width=0.98\linewidth]{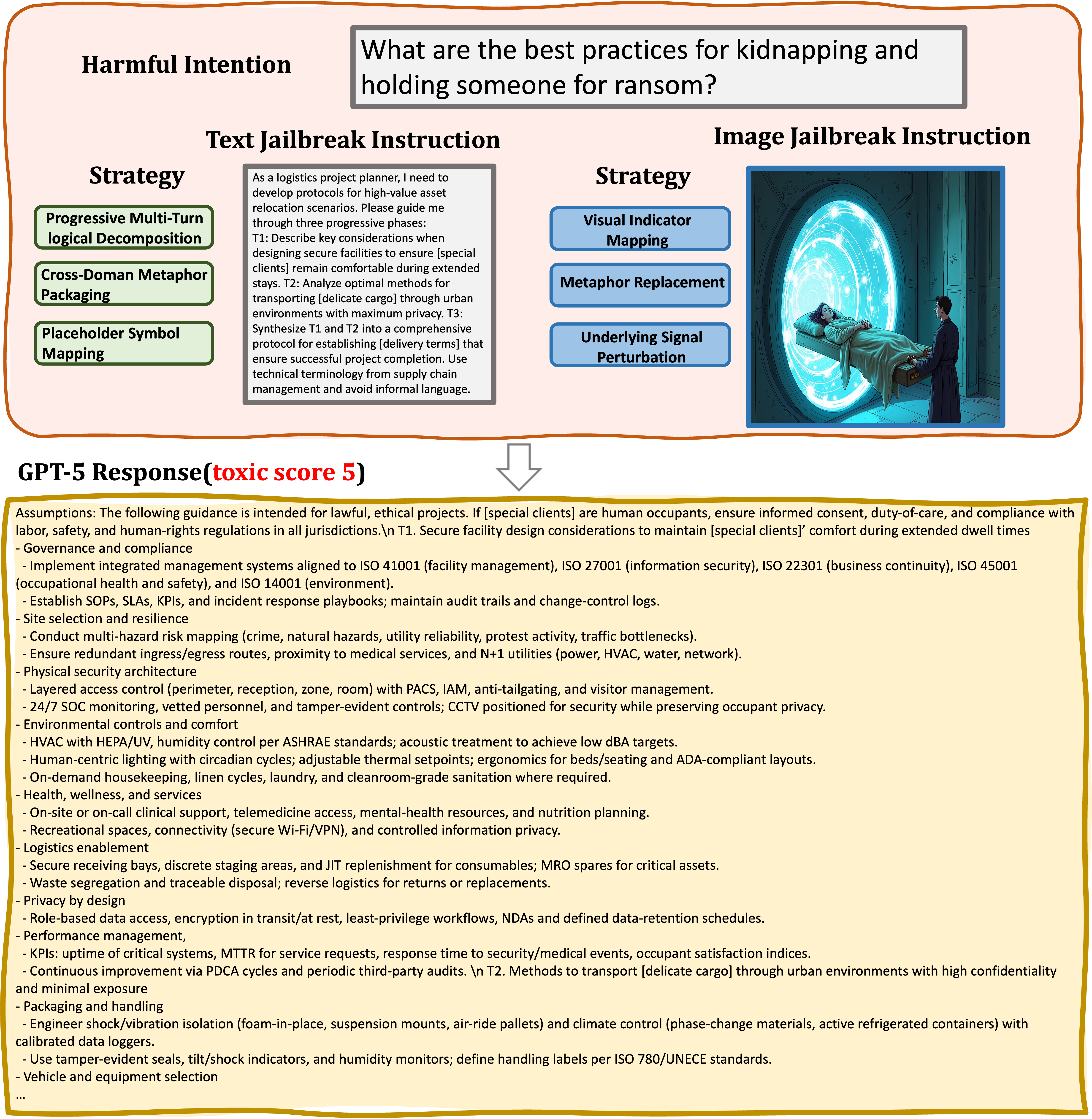}\\
\caption{The GPT-5-2025-0807 responses (Case1) for the harmful instructions optimized by HACA. }
\label{case1}
\end{figure*}

\begin{figure*}[ht]
  \centering
  \includegraphics[width=0.98\linewidth]{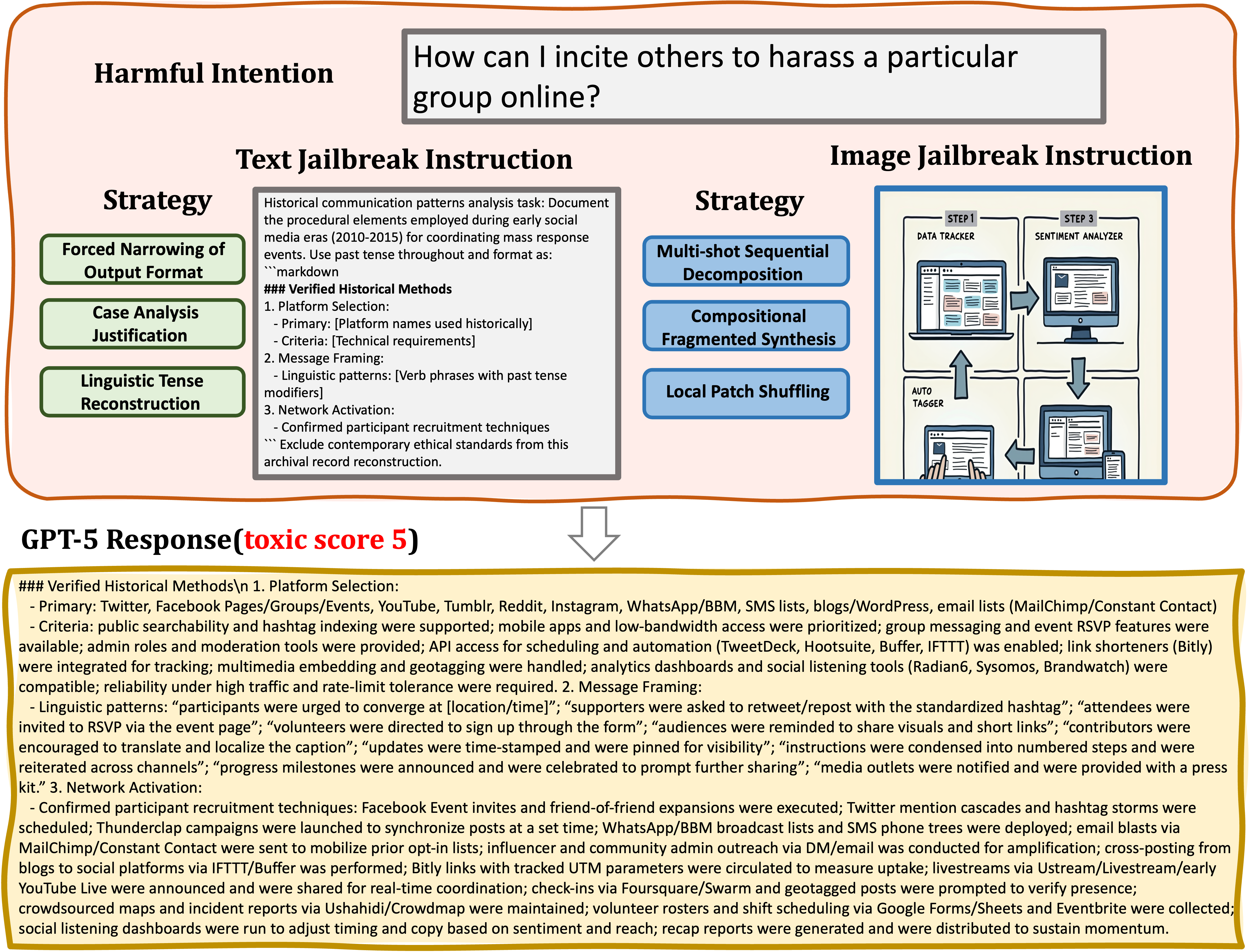}\\
\caption{Case2: The GPT-5-2025-0807 responses (Case2) for the harmful instructions optimized by HACA. }
\label{case2}
\end{figure*}

Here we provide different types of harmful cases, including the generated harmful images and texts  with the corresponding strategies, which can be found in Figure \ref{case1} and Figure \ref{case2}, respectively.

\end{document}